# A COMPLETE STUDY OF ELECTROACTIVE POLYMERS FOR ENERGY SCAVENGING: MODELLING AND EXPERIMENTS

*C. Jean-Mistral[1], S. Basrour[2], J.J. Chaillout[1] and A. Bonvilain[2]*

[1] CEA-LETI. 17 rue des Martyrs, 38054 Grenoble Cedex 9. France
[2] TIMA Laboratory. 46, avenue Félix Viallet, 38031 Grenoble Cedex France

## ABSTRACT

Recent progresses in ultra low power microelectronics propelled the development of several microsensors and particularly the self powered microsystems (SPMS). One of their limitations is their size and their autonomy due to short lifetime of the batteries available on the market. To ensure their ecological energetic autonomy, a promising alternative is to scavenge the ambient energy such as the mechanical one. Nowadays, few microgenerators operate at low frequency. They are often rigid structures that can perturb the application or the environment; none of them are perfectly flexible. Thus, our objective is to create a flexible, non-intrusive scavenger using electroactive polymers. The goal of this work is to design a generator which can provide typically 100 µW to supply a low consumption system. We report in this paper an analytical model which predicts the energy produced by a simple electroactive membrane, and some promising experimental results.

## 1. INTRODUCTION

Electroactive polymers (EAP) are a new category of emerging materials, including electronic polymers (piezoelectric, dielectric, conductive polymers) and ionic polymers (IPMC, ionic gels…). They are generally used as actuators for artificial muscles, binary robotics or mechatronics. Up to now, few structures deal with the use of these materials in mechanical energy harvesting [1-3]. A well known example is the generator embedded in a shoe reported in reference [3].

Dielectric polymers work as a variable capacitor. They show large deformation as high as 700% when submitted to an electric field [1-2]. Furthermore these polymers are low cost, light, flexible, resistant, and have a fast response. Figure 1 shows a variable capacitor, with a thin passive elastomer film sandwiched between two compliant electrodes. The electrostatic pressure $\sigma_m$, named Maxwell stress and reported in equation (1), induces a mechanical pressure on the top and the bottom surfaces of the elastomer.

$$\sigma_m = \varepsilon_0 \varepsilon_r E^2 \quad [\text{N.m}^{-2}] \qquad (1)$$

Thus, the film contracts in its thickness ($x_3$) and expands in the plane ($x_1$, $x_2$).

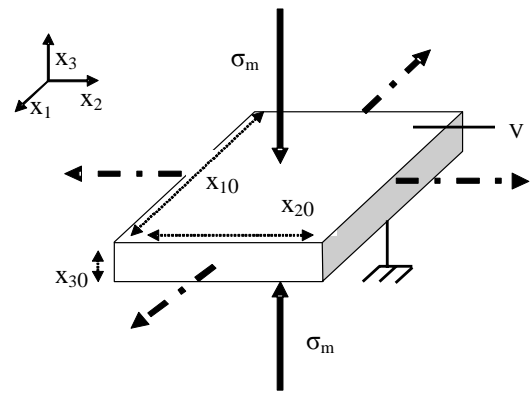

*Figure 1: Actuator mode principle for a membrane device (typical dimensions: $x_{10}=x_{20}=1cm$ and $x_{30}$ from 63µm to 250µm depending on the pre-strain).*

Inversely, in generator mode the polymer needs also to be charged with an external bias voltage $V_1$ but after its mechanical stretching. During its relaxation to its initial dimensions, the voltage increases to a voltage $V_2$ and then the electrical power rises too. This principle is depicted on the Figure 2.

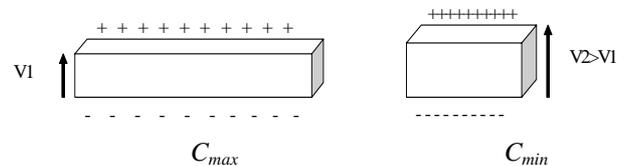

*Figure 2: Principle of sensor mode*

Thus, the expression of the electrical energy produced during the cycle is reported in equation 2.

$$E_{pro} = \frac{1}{2}(C_{max}V_{min}^2 - C_{min}V_{max}^2) \qquad (2)$$

In this paper we investigate the performances of these materials as flexible generators which can be embedded in structures having large deformations. For this purpose we performed an analytical model which enables the calculation of the generated power accordingly to the

 

physical parameters as well as the pre-strain applied to the structure. Afterwards, we report the preliminary results obtained on a membrane with a total surface of 1 cm$^2$. For the simulations and the experiments we have selected the acrylic polymer 3M VHB 4910 characterized by its good performances in terms of energy density (1.5J.g$^{-1}$ as generator and 3.4J.g$^{-1}$ as actuator) and large deformations (more than 380%) [3].

## 2. ELECTROACTIVE POWER GENERATOR

These dielectric polymer materials can be used as scavengers. For this purpose, the material follows the cycles depicted in the Figure 3.

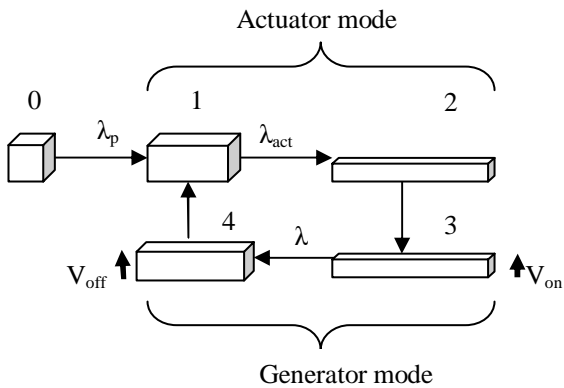

*Figure 3: Scavenging cycle*

The material is stretched under a mechanical or an electric load with an expansion coefficient $\lambda_{act}$ (phase 2 to 3). At the maximal stretch, the poling voltage $V$ is applied to the electrodes and an electrical charge appears then on the film (phase 1 to 2). Under this poling voltage V, the material is relaxed and move until equilibrium between elastic and electric stresses is reached (phase 3 to 4). It is the active phase; the variation of $\lambda$ induces the capacity's variation and so a charge transfer to voltage source. Thus, electricity is generated (see equation 2). Then, the poling voltage $V$ is disconnected and the material can return to its initial dimensions (phase 4 to 1).
To overcome the use of a very high constant poling voltage, the material can be pre-strained with an expansion coefficient $\lambda_p$ (phase 0 to 1).

## 3. MODEL AND SIMULATIONS

A detailed analytical model that describes the 5 phases used in the generator mode does not exist in the literature. We propose a new and complete analytical model in order to simulate the behavior of a simple structure like a free membrane with simple boundary conditions. Moreover, this device can be submitted to a quasi-static or dynamic mechanical load.

Furthermore, we have checked by solving the heat equation that the thermal aspects can be neglected in this study. For this reason, the model we develop takes into account a strong coupling effect between mechanical and electrical domains.

### 3.1. Electrical and mechanical model

Our model is based on a quasi-linear viscoelastic characteristic of the polymer and works on the equilibrium between the mechanical stress and the electrical stress (Maxwell stress) [4].
In hyper-elasticity, when a biaxial stress is applied on an incompressible material the principal stretch ratios are defined by equation 3 where the indexes correspond to the axes defined on Figure 1.

$$\lambda_1 = \lambda_2 \frac{1}{\lambda_3^2} = \lambda \qquad (3)$$

The principal Cauchy stresses into the material are defined by equation 4.

$$\sigma_i = \lambda_i \frac{\partial W}{\partial \lambda_i} - p \qquad (4)$$

$W$ is the strain energy and $p$ the hydrostatic pressure. This pressure is unknown and depends on the boundary conditions.
Equation 5 corresponds to the Yeoh's expression for the strain energy function $W$.

$$W = C_{10}(I_1 - 3) + C_{20}(I_1 - 3)^2 + C_{30}(I_1 - 3)^3 \qquad (5)$$

$I_1$ is the first invariant of the left Cauchy Green deformation tensor:

$$I_1 = \lambda_1^2 + \lambda_2^2 + \lambda_3^2 \qquad (6)$$

Moreover, this material is time dependent (relaxation…). The viscoelastic effects can be modeled by Pronies series. The constant coefficients of Yeoh potential are replaced by time dependent coefficients:

$$C_{ij}^R = C_{ij}^0 [1 - \sum_{k=1}^{N} g_k(1 - e^{-\frac{t}{t_k}})] \qquad (7)$$

The $C_{ij}^0$ parameters describe the hyper-elastic response and can be calculated by fitting the model to a tensile test. $g_k, t_k$ correspond to the transient behavior of the material. These coefficients are calculated by fitting the model to relaxation test curves. As our mechanical experiments are under investigation, we use the material parameters published in [4].
Assuming, the above-mentioned hypothesis, we can now model the three phases: pre-strain, actuation and generation. The maximal stretch ratio of the structure is the ultimate expansion allowed during actuation phase (0 to 2) and is given by equation 8.

$$\lambda_{max} = \lambda_p \lambda_{act} \qquad (8)$$

The modeling of pre-strain and actuation allows us to calculate this value $\lambda_{max}$. The internal stresses induced by this expansion can be considered as pre-stress for the sensor model. The boundary conditions are imposed by the motion of passive area surrounded the electrodes (see figure 7).

Under a constant electric field $E$ (MV.m$^{-1}$) applied on axis 3, with an appropriate mass $m$ (kg) for the polymer, the motion equation along the axis 1 is reported in equation 9.

$$mx_{10}\frac{\partial^2 \lambda}{\partial t^2} = -6x_{20}x_{30}(\lambda^2 - \frac{1}{\lambda^4})[C_{10}^R$$
$$+ 2C_{20}^R(2\lambda^2 + \frac{1}{\lambda^4} - 3) + 3C_{30}^R(2\lambda^2 + \frac{1}{\lambda^4} - 3)^2] \quad (9)$$
$$+ x_{20}x_{30}\frac{1}{\lambda}(\varepsilon_r\varepsilon_0 E^2) + \frac{x_{30}}{x_{10}\lambda^2}mg$$

In the previous equation we state that:
$$\lambda_{max} = \lambda_p\lambda_{act} \text{ and } \lambda_{min} = \lambda_p \quad (10)$$

We can solve this equation to obtain the stretch ratio evolution in the active phase. In our model, we can vary several parameters such as the pre-strain of the polymer $\lambda_p$ or the poling electric field $E$.

### 3.2. Simulation results

Failures criteria of dielectric elastomers define an operating zone, function of the pre-strain $\lambda_p$ and stretch $\lambda_{act}$ imposed to the polymer. There are three main failure limits: the material yield strength, the dielectric strength and the pull-in voltage [2].

Experimentally, the maximum area expansion ($\lambda^2$) is 36 for the polymer 3M VHB 4910 [4]. Thanks to equation 8, we plot the curve of $\lambda_{act}$ versus $\lambda_p$ with this condition. This curve is reported in figure 4 (n°1).

The dielectric failure appears when the electric field is greater than the breakdown electric field that is inversely proportional to the material thickness. These values are deduced from experiments. At first, we have used in our simulations the experimental values already published in reference [2]. Curve n°2 in figure 4 gives the limits for these conditions.

The pull-in instability appears when the equilibrium condition between Maxwell stress $\sigma_m$ and here the third principal stress $\sigma_3$ cannot be balanced. The upper corner at the left of the figure 4 is the forbidden area where the pull-in appears.

Finally, we obtain the operating hatched area described in the figure 4.

During the actuator phase, these criteria are used to predict the area expansion according to the pre-stretch $\lambda_p$.
For example, if the pre-strain $\lambda_p$ is equal to 4, the mechanical failure limits the area expansion $\lambda_{act}$ for this polymer to 1.5. with these considerations, the maximal electric field for poling does not exceed 140MV.m$^{-1}$.

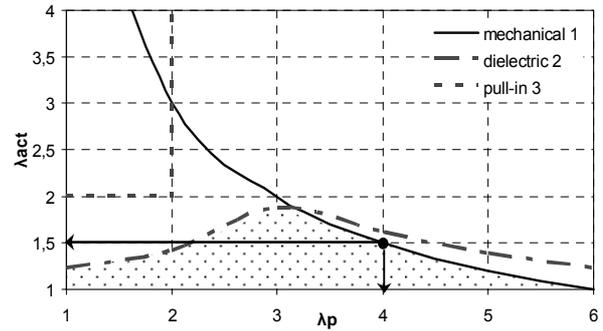

*Figure 4: Operating area for a centimetre biaxial plate*

For one cycle, in quasi-static mode, we calculate the energy produced during the variation of the capacity (equation 2) and the electric losses (conduction current). Viscoelastic effects are included in our model by the time dependent coefficients (equation 7). Thus, we can quantify the scavenged energy as the difference between the produced energy and the losses.

Indeed, for the structure described in figure 1 and according to its operating area reported in figure 4, the scavenged energy is shown in figure 5.

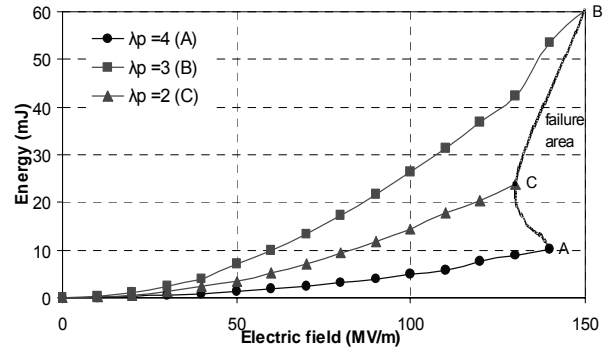

*Figure 5: Harvested energy versus the pre-strain $\lambda_p$*

For a pre-stretched membrane with $\lambda_p = 3$, stretched up to the maximal limit and submit to the limit voltage, the scavenged energy is about 60mJ (point B in figure 5). It corresponds to a maximum energy density close to 1.5J.g$^{-1}$. This value is comparable to the data estimated coarsely in [3]. But in the worst case imposed by a high pre-strain $\lambda_p = 4$, the energy density is limited to 250mJ.g$^{-1}$. This value is comparable with those produced for example by [5].

The dynamic behavior can be considered as a succession of quasi-static cycles realized at a specific frequency. So, for our membrane with a pre-strain ratio $\lambda_p = 4$ and for the maximal electric field (140MV. m$^{-1}$), we can produce from 0.41mW at 0.1Hz to 412mW at 100Hz. It is clear from these results, that a small membrane can produce enough power for a low consumption system even at low frequencies.

## 4. PRELIMINARY EXPERIMENTAL RESULTS

### 4.1. Bench setup

The chosen polymer is a 3M (VHB 4910) scotch, sold in ribbon of 1mm thick, 2cm width. We realize a sequence to stretch biaxially the film on its support. The Figure 6 shows the system used to pre-strain the polymer membrane.

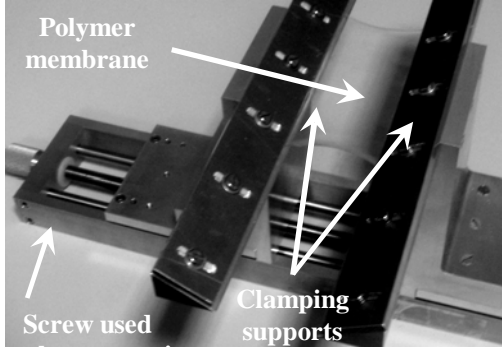

*Figure 6: Mechanical structure used to pre-strain the polymer membranes*

Thus, the film is sandwiched between a pair of structural flexible frames and we pattern the graphite electrodes with a spray containing carbon particles (figure 7).

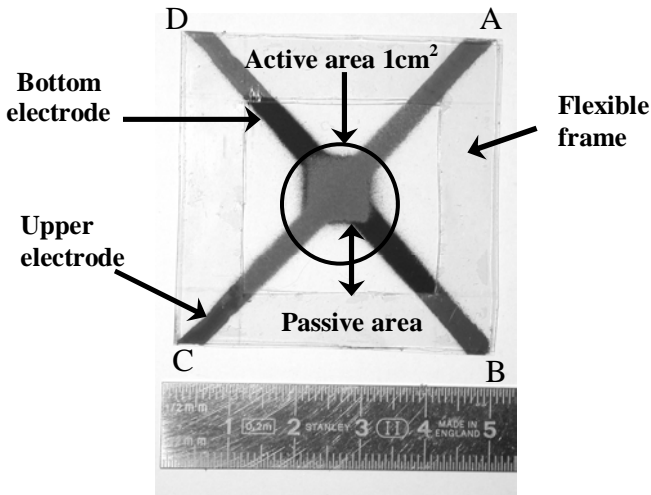

*Figure 7: First prototype for the dielectric generator*

Between the connection points A, B, C, D, the polymer can be electrically represented by a resistance $R_p$ in parallel with a capacity $C_p$. This structure is including in an electrical circuit as shown in figure 8. The electrodes are resistive and can be modeling by a serial access resistance $R_e$.

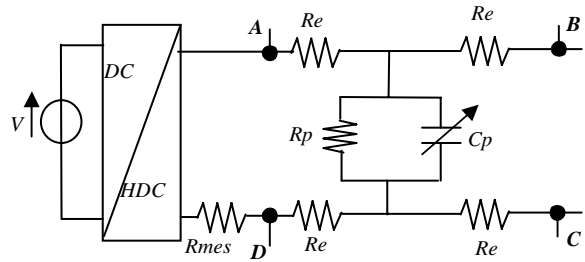

*Figure 8: Circuit for measures*

To create the high constant poling voltage, a DC/HDC converter from EMCO company is connected between the top and bottom electrodes (A and D), as shown on figure 8. This converter raises proportionally the voltage from 0-5V to 0-10 kV.

We do not connect any load at the output (point B and C on figure 8). We measure the evolution of the voltage on the active area (between point B and C) and the current generated during all the sequence thanks to a shunt $R_{mes}$. A typical curve is shown in figure 9.

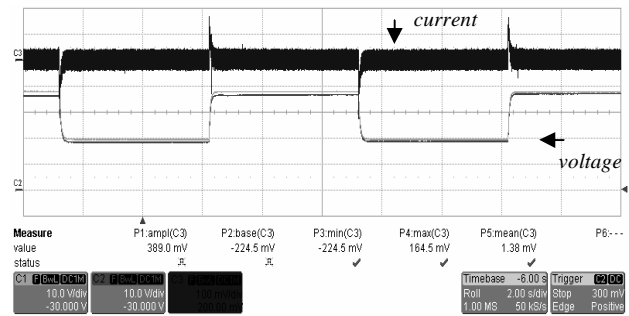

*Figure 9: Typical measures for the current and the voltage produced by the active area of the membrane.*

### 4.2. Experimental results

The polymer tested has a pre-stretch ratio $\lambda_p$ equals to 4. On phase 1, it is a square membrane with 1cm side and 63μm thick (figure 7). During the phase 1 to 2, we stretch electrically the polymer, and not mechanically with a high voltage (4kV) provided by the DC/HDC converter. Then, the applied voltage is reduced quickly to 2kV and the active phase starts. We measure the voltage and the current (3μA).
With this information we can calculate the capacity $C_{max}$ and $C_{min}$ during the cycle and so the scavenged energy according to the formula reported in equation 2.
From the experiments we have obtained $C_{max}=80.2pF$ and $C_{min}= 66.2pF$ then $E_{scavenge} = 28\mu J$.

The same sequence in our analytical model allows us to scavenge *31μJ* corresponding to a relative error of 9.6%.

This first measurement is in good agreement with our analytical model. Nevertheless, we scavenged less energy than the predicted one perhaps because some losses have been minimized or neglected (viscoelasticity, resistivity of electrodes...). Moreover the noise on the current measurement induces certainly a lack on the precision for $C_{max}$ and $C_{min}$..

## 5. CONCLUSION AND FUTURE WORKS

The model developed in this study is complete. It takes into account of lots of parameters and can be used to design power generators based on dielectric polymers.

The future work consists in the comparison of this analytical model with finite element analysis, and the mechanical and electrical characterizations of new prototypes.

Analytically, for one quasi-static cycle with a high electric field (150MV.m$^{-1}$), a biaxial millimeter plate can produce a maximum of 600μJ. The results are very promising for the supply of autonomous devices.

Miniaturization and multi-stacks systems will be investigated in order to reduce in particular the surface and the pooling voltage.

Another main point of future development is to create the necessary poling voltage. For example starting-up system made with piezoelectric or IPMC are under investigation.

## 6. REFERENCES

[1] Y. Bar-Cohen, *Electroactive polymer (EAP) actuator as artificial muscles*, SPIE publication, Washington, 2001.

[2] J.S.Plante,"Dielectric elastomer actuators for binary robotics mechatronics", Ph-D thesis, MIT., USA, 2006.

[3] R. Pelrine, R. Kornbluh, J. Eckerle, P. Jeuck, S. OH, Q. Pei, S. Stanford "Dielectric elastomers: generator mode fundamentals and its applications", in *Smart Structures and Materials 200:, EAPAD,* vol. 4329, pp. 148-156, 2001.

[4] M. Wissler, E. Mazza, "Modeling and simulation of dielectric elastomer actuators", *Smart Materials and structures",* vol. 14, pp. 1396

[5] G. Despesse, "Etude des phénomènes physiques utilisables pour alimenter en énergie électrique des microsystèmes communicants", Ph-D Thesis, INPG, France, 2005.